\documentclass[11pt]{article} 

\usepackage{arxiv}
\usepackage[hyphens]{url}

\usepackage[utf8]{inputenc} 
\usepackage[T1]{fontenc} 
\usepackage{mathpazo} 
\usepackage{ragged2e} 
\usepackage{float, graphicx}
\usepackage{indentfirst} 
\usepackage{titlesec}
\usepackage{amsthm, amssymb, mathtools} 
\usepackage{caption}
\usepackage{subcaption}

\usepackage{centernot}

\usepackage{tikz}

\usepackage[breaklinks]{hyperref}

\usepackage{wrapfig}
\usepackage{bm}
\usepackage{lipsum}

\usepackage{longtable,booktabs}

\usepackage{comment}

\newcommand{\supplementarysection}{%
  \setcounter{figure}{0}
  \let\oldthefigure\thefigure
  \renewcommand{\thefigure}{S\oldthefigure}
}

\title{An informational approach to uncover the age group interactions in epidemic spreading from macro analysis}

\author{{Tiago Martinelli} \\ 
    University of S\~ao Paulo \\
	São Carlos, Brazil\\
	\texttt{tiago.martinelli93@gmail.com} \\
	\And
	{\hspace{1mm}Alberto Aleta} \\
	Institute BIFI and Dept. of Theoretical Physics\\
	University of Zaragoza, Spain \\
	\texttt{albertoaleta@gmail.com} \\
	\AND
	{Francisco A. Rodrigues} \\
	University of S\~ao Paulo \\
	São Carlos, Brazil \\
	\texttt{francisco@icmc.usp.br} \\
	\And
	{\hspace{1mm}Yamir Moreno} \\
	Institute BIFI and Dept. of Theoretical Physics\\
	University of Zaragoza, Spain \\
	CENTAI, Turin, Italy\\
	\texttt{yamir.moreno@gmail.com} \\
}	

\begin{document}
\maketitle

\begin{abstract}

We investigate the use of transfer entropy (TE) as a proxy to detect the contact patterns of the population in epidemic processes. We first apply the measure to a classical age-stratified SIR model and observe that the recovered patterns are consistent with the age-mixing matrix that encodes the interaction of the population. We then apply the TE analysis to real data from the COVID-19 pandemic in Spain and show that it can provide information on how the behavior of individuals changed through time. We also demonstrate how the underlying dynamics of the process allow us to build a coarse-grained representation of the time series that provides more information than raw time series. The macro-level representation is a more effective scale for analysis, which is an interesting result within the context of causal analysis across different scales. These results open the path for more research on the potential use of informational approaches to extract retrospective information on how individuals change and adapt their behavior during a pandemic, which is essential for devising adequate strategies for an efficient control of the spreading.

\end{abstract}

\section{Introduction}

Mathematical epidemiology models are crucial for understanding the spread of infectious diseases, but they must be refined to incorporate the heterogeneities of the population. This includes factors such as the age of individuals, their socioeconomic status, mobility patterns, and even behavioral changes \cite{Aleta2020Jul}. However, these approaches can be quite demanding in terms of data, and the necessary data are often difficult and expensive to collect. For example, although epidemiological models began incorporating age as a factor as early as the beginning of the twentieth century \cite{MKendrick1925Feb}, it was not until 2008 that the POLYMOD collaboration published the first large-scale survey to measure age-based contact patterns in the population \cite{Mossong2008Mar}. These surveys enable the construction of age contact matrices that encode the number of contacts between individuals of different age groups. However, these matrices are highly dependent on demographic, sociological, and cultural patterns of the population, making extrapolation difficult.

Although measuring age contact patterns through surveys is the most direct method, it is often slow, expensive, and subject to biases due to self-reporting by survey participants \cite{Hoang2019Sep}. An alternative approach is to use public socioeconomic data to infer these patterns, which tipically involves dividing interactions into four settings: schools, households, workplaces, and the community \cite{Mistry2021Jan}. The number of contacts in schools and households can be estimated relatively easily using available information on their size. In contrast, workplace contacts are more difficult to determine, as only aggregated statistics are typically available. Finally, community contacts encompass all other types of interaction, such as those in public transportation, shopping, social encounters, or any other random contacts. These contacts are often approximated or extrapolated from surveys carried out in similar regions \cite{Prem2021Jul}.

However, it is important to note that these methods only provide a snapshot of the contact patterns of the population. Due to the dependence of these patterns on the demographic structure of the population, they are not static and should reflect population changes over time \cite{Arregui2018Dec}. Additionally, these methods assume a baseline state where individual behavior remains constant. However, the recent COVID-19 epidemic has demonstrated the significant impact that epidemics can have on human behavior, both through the propagation of the disease and as a result of non-pharmaceutical interventions implemented to control it \cite{Hunter2021Jun,Aleta2022Jun,Klein2022Dec}. Although surveys conducted during the pandemic period have shown significant changes in the contact patterns of the population, they are limited to specific places and times \cite{Zhang2020Jun,Koltai2022Mar,Feehan2021Feb}.

It is thus of paramount importance to devise methods that can provide retrospectively information on how individuals change and adapt their behavior during a pandemic, as it can have a huge impact on epidemic outcomes \cite{Pollan2020Aug}. In this work, we propose an informational approach to unfold these changes in the contact patterns of the population from commonly collected surveillance data. Due to privacy concerns, publicly available data is mostly restricted to new cases (incidence), hospitalization rates and deaths. Besides, these are seldom stratified beyond large regions or age groups. As such, we focus on studying the variations on the age contact patterns of the population and propose an informational approach based on transfer entropy (TE) quantifiers that can unfold these changes from incidence curves. Furthermore, our results show that prior knowledge on the dynamical process that created the data provide better estimates for the observable in question by coarse-graining the data. Since the methodology presented in this paper can be easily generalized to other processes beyond the context of epidemic spreading, these results also open the path for new applications of coarse-graining techniques in information theory.

\section{Epidemic dynamics}
\begin{figure}
\begin{center}
\includegraphics[width=\textwidth]{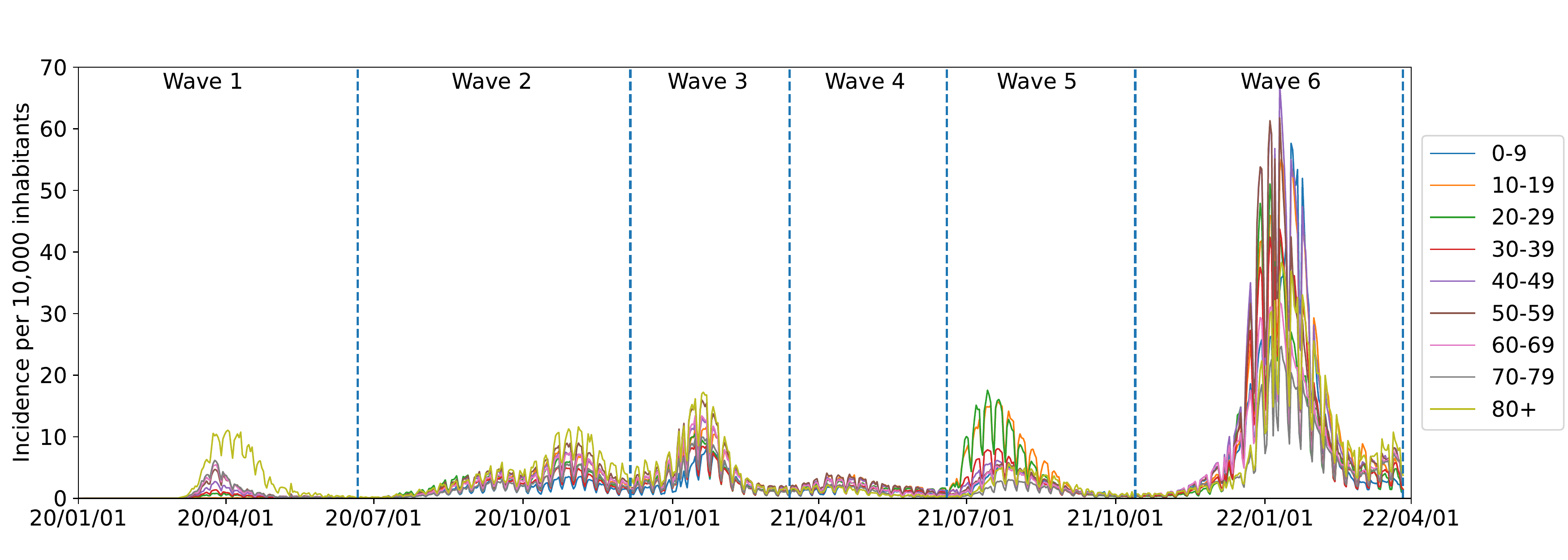}
\caption{Incidence (new cases) per 10,000 inhabitants in Spain. Each curve represents the incidence for a certain age group, with 10 years wide brackets from 0-9 to 70-79 and one last bracket for people 80 years old or older. The Spanish government classified the propagation in 6 waves with breakpoints: [2020-01-01, 2020-06-21]; [2020-06-21, 2020-12-06]; [2020-12-06, 2021-03-14]; [2021-03-14,2021-06-19]; [2021-06-19, 2021-10-13], [2021-10-13, 2022-04-01] \cite{COVIDdata, Informe162}.}
\label{fig:evolution}
\end{center}
\end{figure}

Our data represents the incidence per 10,000 inhabitants of COVID-19 cases recorded from 2020 to 2022 in Spain \cite{COVIDdata}. The data is segmented by age groups, with each group spanning a 10-year range, from 0-9 to 70-79, and one last group for people 80 years old or older (9 groups). Further, the Spanish government classified the propagation in the 6 waves shown in figure \ref{fig:evolution} (note that the surveillance system stopped collecting data on a consistent basis after the 6th wave). We exclude the first wave from our analysis due to limitations in the surveillance system during that time period, as evidenced by seroprevalence studies indicating that the number of actual infections was likely much higher than the reported cases, and these were biased towards older age groups as they were more likely to require hospitalization \cite{Pollan2020Aug}.

The simplest model that can illustrate the impact of age-mixing patterns on the spread of COVID-19 is the age-stratified susceptible-infected-recovered (SIR) model \cite{Zhang2020Jun}. This model stratifies the population $N$ into several age groups that interact with each other according to the age-mixing matrix $M_{aa'}$. This matrix encodes the average number of contacts that an individual in age group $a$ has with individuals in age group $a'$ (see figure \ref{fig:matrix}). Within these age groups, individuals are further classified into three groups: susceptible (S), infected (I) or recovered (R). Infected individuals may infect the susceptible population with rate $\beta$, and recover with rate $\gamma$, as shown in equations \eqref{meso_eq1}, \eqref{meso_eq2} and \eqref{meso_eq3}. Note that if we set $M_{aa'} = N_{a'}/N$, where $N_{a'}$ is the size of age group $a'$ and $N$ is the total population size, we would recover the classical homogeneous mixing SIR model.

\begin{center}
\begin{minipage}{.45\textwidth}
\begin{eqnarray}
\dot{S}_a &=& - S_a\sum_{a'} \beta M_{aa'}\frac{I_{a'}}{N_{a'}}  \label{meso_eq1}\\ 
\dot{I}_a &=&  S_a\sum_{a'} \beta M_{aa'}\frac{I_{a'}}{N_{a'}} - \gamma I_a \label{meso_eq2}\\ 
\dot{R}_a &=&  \gamma I_{a'} \label{meso_eq3}
\end{eqnarray}
where $N_{a'}$ is the number of individuals in age group $a'$ and $N = \sum_{a} N_{a} = \sum_{a} S_a + I_a + R_a$. 
\end{minipage}%
\hspace{.65cm}
  \begin{minipage}{.45\textwidth}
\begin{figure}[H]
    \centering
    \includegraphics[width=7cm,height=6cm]{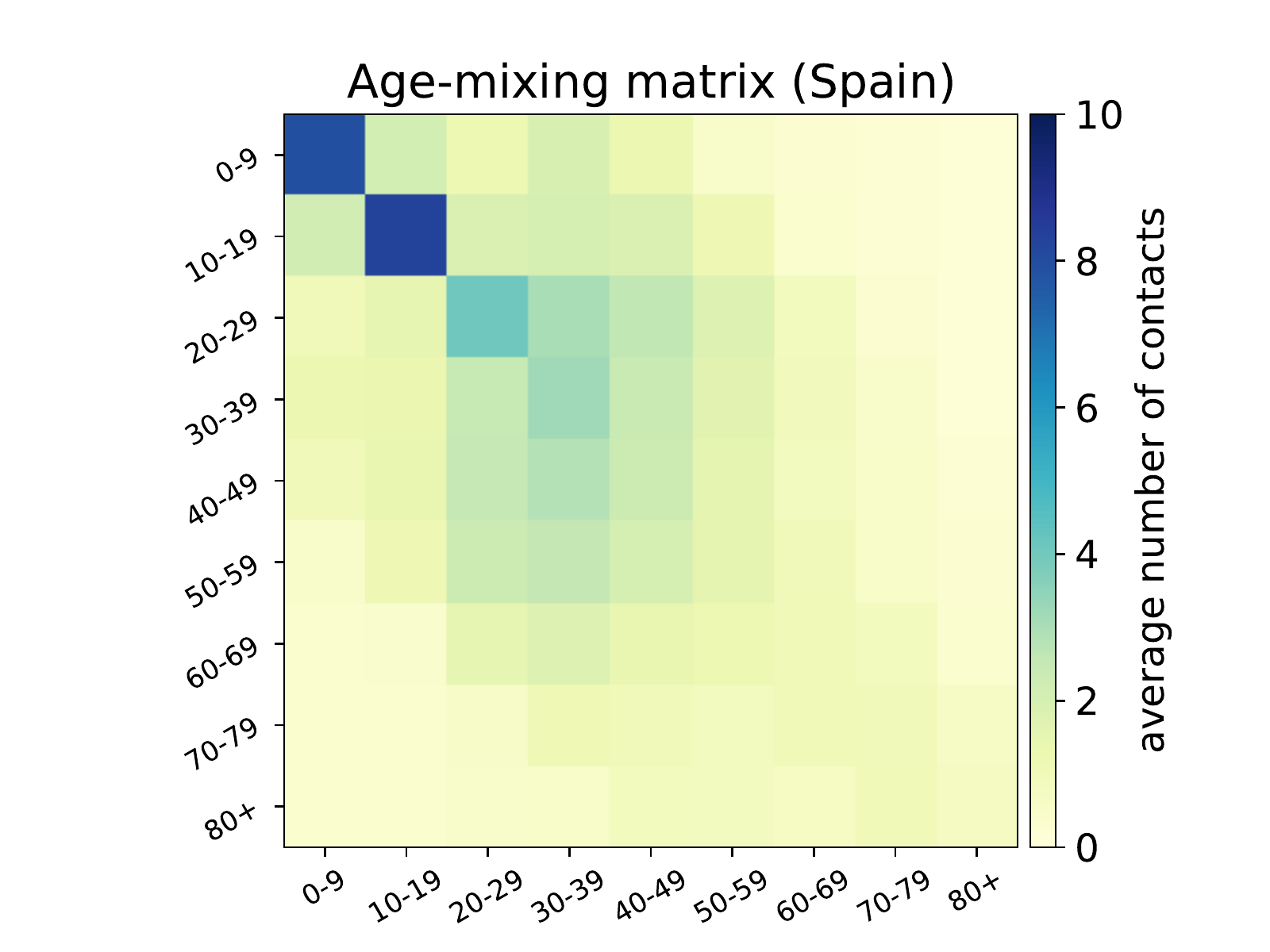}
    \captionsetup{width=.75\textwidth}
    \vspace{-.25cm}
    \caption{\small{Estimated age-mixing matrix using demographic and census data from Spain \cite{Mistry2021Jan}.}}
    \label{fig:matrix}
\end{figure} 
\end{minipage}    
\end{center}

For our analysis, we will consider a multivariate time series $\bm{X}^{V} = (X^{V}_t)_{t\in \mathbb{N}}$ with the spatial index $V$ being the vector of different age groups represented by the different colors in Fig.\ref{fig:evolution}. These time series represent the incidence per age group which corresponds to $I_a/N_a$ in the SIR model. To study the dynamic relationships over time among the series, we specified them in terms of the conditional distribution of $X_t$ given its past up to order $\tau$, denoted by $X_{t}^{(\tau)} = (X_{t-s})_{s = 1}^{\tau}$. Thus, we will describe our time series data as a family of probability density kernels $p$ from $\mathbb{R}^{V\times \mathbb{N}}$ to $\mathbb{R}^{V}$ \cite{book_ts1}.

To optimize the capture of the past influence of the time series we can leverage the dynamics of the transmission process. As individuals become infectious, they may propagate the virus to other susceptible individuals until their infectivity vanishes. Thus, the number of new cases at time $t$ will be approximately proportional to the number of individuals that are still infectious at that time:

\begin{equation}
    \label{eq:gt}
    C(t) \propto \sum_{s=1}^t \phi(s) C(t-s)\,
\end{equation}
where $C(t)$ is the number of cases at time $t$ and $\phi(s)$ is the generation time \cite{Zhang2020Jul}. The generation time is defined as the probability distribution function for the time from infection of an individual to the infection of a secondary case by that individual. In the SIR model the generation time is given by an exponential distribution with rate $1/\gamma$ \cite{Wallinga2006Nov}. In the case of the COVID-19 epidemic the generation time can be approximated by a gamma distribution with shape 1.87 and rate 0.27 \cite{Cereda2021Dec,Manica2023,Manica2022Aug}.

\section{Information-theoretic analysis}

Our analysis relies on measuring the interaction between different age groups using informational metrics, namely, transfer entropy, that shares some of the desired properties of mutual information but takes the dynamics of information transport into account \cite{TE}. Transfer-like entropies (TEs) have become a widely accepted measure of information transfer \cite{book_transfer}. They are defined as the mutual information between the past of source(s) time-series process(es) and the present state of target process(es), conditioned on the past of the target. Mathematically (in discrete time), is written as:

\begin{equation}
T_{X\rightarrow Y}(k,l) := I(X_{t}^{(k)} ; Y_{t}| Y_{t}^{(l)}) = 
\mathbb{E}_{p(x_{t}^{(k)},y_{t}^{(l)}, y_{t})}\Bigg[-\log
\frac{ p(x_{t}^{(k)},y_{t}|y_{t}^{(l)})}{p(x_{t}^{(k)}|y_{t}^{(l)}) p(y_{t} |y_{t}^{(l)})}\Bigg].
\label{eq:te}
\end{equation}Here, $X^{(k)}_t$ and $Y_{t}^{(l)}$ are random vectors representing the past influence of the processes $X_t$ and $Y_{t}$ up to order $k$ and $l$, respectively; the letter $I$ represents the conditional mutual information \cite{CoverThomas}.

It is also possible to condition the TE on additional processes, getting measures as multivariate transfer entropy \cite{2015_Bollt, novelli}.  However, we are considering that the different age groups' influence is uncorrelated to each other in the moment of the influence. When combined with a suitable statistical significance test, the TE can be used to show that the present state of $Y_t$ is conditionally independent of the past of $X_t$ when conditioned on the past of $Y_t$. Here, we refer to conditional independence in the statistical sense, i.e., $p(x^{(k)},y_{t}|y^{(l)})=p(y_{t}|y^{(l)})$.

\subsection{CMI estimator for continuous data}

Due to the continuous nature of our data the idea here is to directly estimate CMI by using the well-established Kozachenko-Leonenko k-nearest neighbor estimator \cite{frenzelpompe, kl},
\begin{equation}
\widehat{I}(X;Y|Z) = \frac{1}{N} \sum_{i=1}^N
\left[\psi(k_{XZ,i}) + \psi(k_{YZ,i}) - \psi(k_{Z,i})\right] - \psi(k)   
\end{equation}where $\psi$ is the digamma function defined as the logarithmic derivative of the gamma function $\psi(x) = \frac{d}{dx}\log \Gamma(x)$ and sample length. The unique free parameter, the number of nearest-neighbors $k$, can be viewed as a density smoothing parameter. For large $k$, the underlying dependencies are more smoothed and CMI has a larger bias but lower variance, which is more important for significance testing. The decisive advantage of this estimator compared to fixed global bandwidth approaches is its local data-adaptiveness, see App.\ref{app:knn} for further details. We compute the knn-estimator using the Python library from ref.\cite{Wollstadt2019}.

\subsection{Statistical Test}

A crucial step in the TE analysis relies on conditional independence tests (CITs) \cite{statistical_tests}, i.e., determining whether the CMI is positive or not. Due to the finite sample size, the CMI estimators may produce nonzero values in the case of zero CMI, and it may even return negative values if the estimator bias is larger than the true CMI \cite{roulston, ksg}. Since no theory on finite sample behavior of the CMI estimator is available, we will use a permutation-based generation of the distribution under the null hypothesis \cite{statistical_tests}. In the absence of an analytic solution, the null distributions are computed in a nonparametric way by using surrogate time series \cite{SCHREIBER}. The surrogates are generated to satisfy the null hypothesis by destroying the temporal relationship between the source and the target while preserving the temporal dependencies within the sources \cite{ete}, see App.\ref{app:shuff} for further details.

\subsection{Coarse-grainning past influence}

As previously described, to facilitate the detection of the causal relationships between the incidence of different age groups, we coarse-grain the data using the generation time distribution. To optimize the capture from the past influence of the time series we coarse-grained it weighted by a gamma distribution in the respective past window. Temporal coarse-graining can be seen as an interventional procedure, see App.\ref{app:eff}. When the intervention strategy is well-defined, this approach can eliminate micro-level confounders as well as problems associated with selection bias.  

The maximal lag $\tau$, the length of the time series window, is particular to the kind of data. In our case, this value can be potentially infinite, since any individual who was infected at a past point may still be infected. For practical purposes, we set $\tau=10$ days, for which the CDF of the Gamma distribution is over 0.8. Then, we have the following mapping:
\begin{equation}\label{map}
\mathbf{X}_{t}^{(\tau)} \mapsto X_{agg}^{(\tau)} := \bm{\gamma}_{\tau}^\intercal\mathbf{X}_{t}^{(\tau)}, 
\end{equation}where each term above represents, respectively, the random vector $\mathbf{X}_{t}^{(\tau)} := (X_{t-1}, X_{t-2}, \ldots, X_{t-\tau})$ and the transposed gamma vector generated as described before. With this, we aggregated the past influence in the unique random variable $X_{\text{agg}}^{(\tau)}$.

\section{Results \label{results}}

\begin{figure}[h]
\begin{center}
\includegraphics[width=\textwidth]{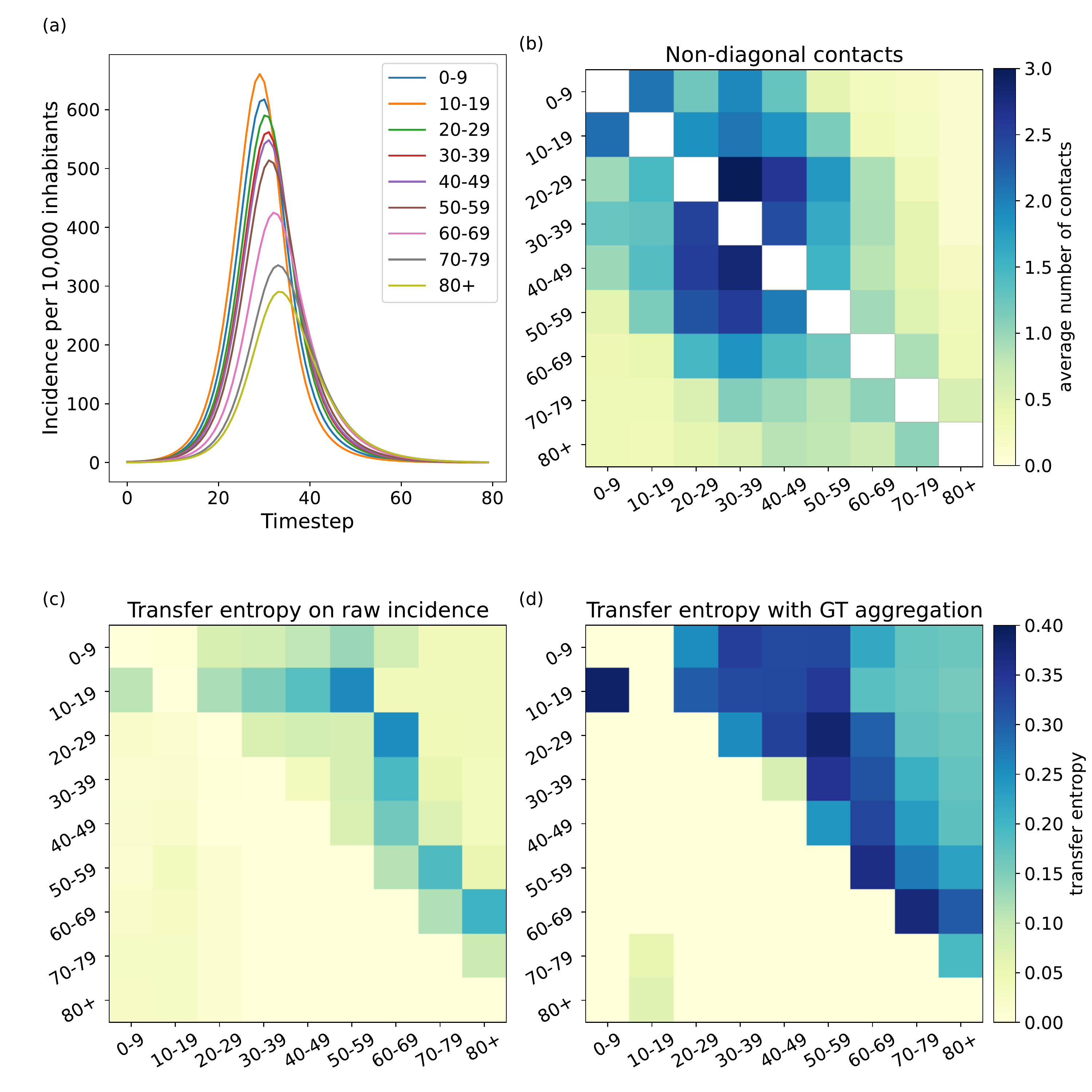}
\vspace{-.25cm}
\captionsetup{width=.95\textwidth}
\caption{Transfer entropy analysis of an age-stratified SIR model. (a) Incidence per age group with $\beta=0.035$ and $\gamma = 1 / 5$. (b) Age-mixing matrix without the diagonal contacts to highlight the interactions across groups. (c) Transfer entropy obtained using the raw incidence. (d) Transfer entropy obtained with the aggregated time series over the generation time.}
\label{fig:te_model}
\end{center}
\end{figure}

We first look at the signal of the contact patterns that can be extracted from the TE analysis of time series data. Figure \ref{fig:te_model}a shows the incidence obtained from equations \eqref{meso_eq1}-\eqref{meso_eq3} and the age-mixing matrix displayed in figure \ref{fig:matrix}, which uses common spreading parameters. Since the TE cannot measure a time series' interaction with itself, figure \ref{fig:te_model}b highlights the non-diagonal contribution of the age contact patterns. Figure \ref{fig:te_model}c displays the TE estimated using the raw incidence curves. We observe that individuals in the 10-19 age group made the most significant contribution to the TE, which aligns with their higher incidence rates. Notably, the interaction between highly clustered age groups such as 20-29, 30-39, and 40-49 was not as clearly observable in terms of TE. We also noted that older age groups were primarily infected by groups immediately preceding them, which was reasonable given their low connection rates with the rest of the population. Lastly, in figure \ref{fig:te_model}d, we demonstrate that coarse-graining the same information using an exponential distribution results in a much larger TE signal.

A few observations are in order. First, when we remove the age contact patterns, the transfer entropy across time series is 0 (see figure \ref{fig:model_noage}). This indicates that any TE larger than 0 will be related to the contact patterns of the population. Second, the assumption of independence may hamper the extraction of patterns across highly clustered age groups. Third, with the baseline behavior we observe an almost upper diagonal matrix, indicating that the spreading tends to go from the young to the old population. This is consistent with the shape of the age-mixing matrix, where the age groups with more contacts - and, hence, higher probability of getting infected - are the youngest ones. Then, interactions across groups tend to follow a cascade pattern with younger individuals interacting more with the ones immediately older than them. Thus, any deviations that we may find from this result can be associated with changes in the contact patterns of the population.

A few observations are in order. Firstly, when we remove the age contact patterns, the transfer entropy across time series becomes 0 (see figure \ref{fig:model_noage}). This implies that any TE value greater than 0 will be related to the population's contact patterns. Secondly, the assumption of independence may hinder the extraction of patterns across highly clustered age groups. Thirdly, with baseline behavior, we observed an almost upper diagonal matrix, indicating that spreading tends to move from younger to older populations. This aligns with the age-mixing matrix's structure, where the youngest age groups are that interact the most, and therefore have a higher probability of infection and transmission. Interactions across age groups generally follow a cascade pattern, with younger individuals interacting primarily with those immediately older than themselves. Thus, any deviations we observe from this upper diagonal structure could be associated with changes in the behavior of the population and their contact patterns.

\begin{figure}[h]
\begin{center}
\includegraphics[width=\textwidth]{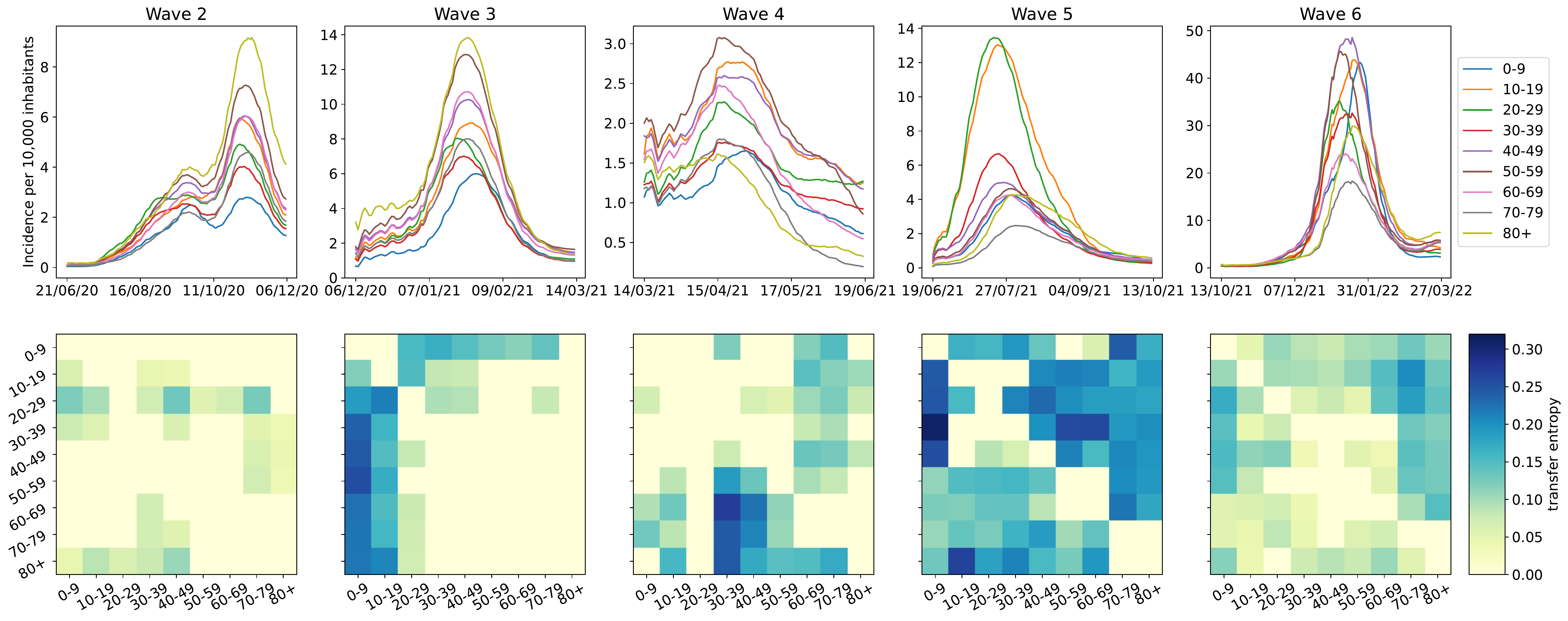}
\vspace{-.25cm}
\captionsetup{width=.95\textwidth}
\caption{Transfer entropy analysis of the COVID-19 pandemic in Spain. Top row: incidence per age group in each official wave. The curves have been smoothed using a rolling average of 14 days to improve visualization. Note also that the scale of the $y$ axis changes in each wave. Bottom row: estimated TE across age groups in each wave.}
\label{fig:te_data}
\end{center}
\end{figure}

Next, we measured the TE from incidence curves collected between 2020 and 2022 in Spain, as shown in figure \ref{fig:te_data} (see figure \ref{fig:S_te_data} for the analysis without coarse-graining the time series). We observe substantial changes in the TE pattern for each wave. Wave 2 was the longest and characterized by the summer period in Spain, as well as the reopening of schools and universities in late September. This might have reduced children's contribution to the spreading process, erasing the baseline pattern. Wave 3 was primarily associated with the Christmas period, during which children tend to be at home and reunite with their families. Wave 4 was the smallest and marked by the vaccination of older age groups. Wave 5 corresponds to the summer of 2021, although the start of the wave was associated to young individuals attending parties to celebrate the end of the academic year \cite{young}. This pattern was closest to the baseline scenario. The final wave was the largest in terms of detected cases, despite having the lowest impact on hospitalization rates and deaths. It was also associated with the Christmas period, although the incidence and TE estimations differed significantly from those of the third wave.

\begin{figure}[h]
\begin{center}
\includegraphics[width=\textwidth]{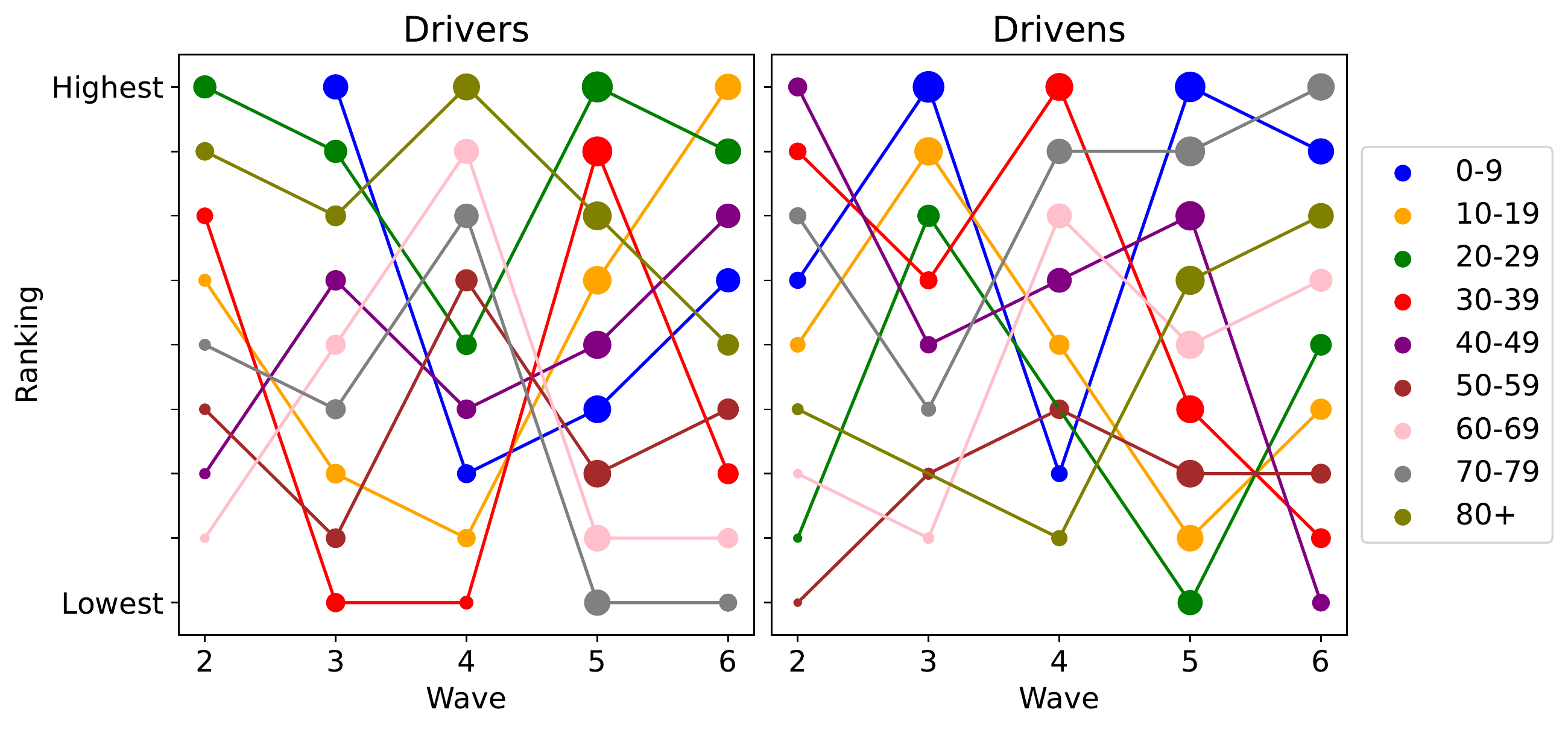}
\vspace{-.25cm}
\captionsetup{width=.95\textwidth}
\caption{Age groups that tend to drive or are driven by others. We sum the total TE associated to each group and rank them according to its value. The size of the markers is proportional to the total TE. Missing markers indicate that no TE was measured for that age group in the given wave.}
\label{fig:driver}
\end{center}
\end{figure}

To provide a more comprehensive view of the changes in the contact patterns of the different age groups, we measure the total TE in each direction. This way, the age group $a$ represented by the time series $X_a$ will be a driver if the value of $T_{X_a} = \sum_{a'} T_{X_a \rightarrow X_{a'}}$ is high, and a driven if $T_{X_a} = \sum_{a'} T_{X_{a'} \rightarrow X_a}$ is high. In figure \ref{fig:driver}, we rank the age groups according to these values as a function of the wave. We observe that people in the 20-29 age group are among the major drivers in many waves except for the 4th one. Note that this wave was the smallest one and also took place while the older age groups were being vaccinated. The age groups that are most commonly driven are those aged 0-9 and 70-79 years old.

\section{Conclusion}

In this study, we aimed to explore the potential of using transfer entropy (TE) to detect the contact patterns of the population during a pandemic. Understanding these patterns is critical for designing effective public health interventions, and properly gauging the potential spread of an epidemic. To achieve this, we first analyzed the results of a classical age-stratified SIR model and found that the TE patterns were consistent with the age-mixing matrix. In particular, we observed that the group 10-19 was the major contributor to the spread of the virus, and that elder age groups tended to be infected by groups situated immediately before them due to their low number of connections with the rest of the population.

We then looked at the interaction between different age groups during the COVID-19 pandemic in Spain. It is worth noting that the age-mixing matrix of Spain was inferred from public data and not directly measured through surveys \cite{Mistry2021Jan} In addition, younger individuals were less susceptible to the infection \cite{RussellMViner2021Feb}. Thus the TE between time series should differ from the baseline model. Nevertheless, the most interesting aspect is to be able to provide a reliable measure of who were the most relevant age groups in each wave and how they changed as the pandemic evolved. These findings provide valuable insights into the transmission dynamics of COVID-19, which can help public health authorities to adapt their response strategies to more effectively control the spread of the virus.

Lastly, we have shown how knowledge of the underlying dynamics of the process allows us to build a coarse-grained representation of the time series that provides more information than the raw time series. Therefore, the macro-level representation of the process is more effective for analysis (see also Appendix \ref{app:eff}). This is an interesting result in the context of causal analysis across different scales. These findings also suggest the potential for using informational approaches to retrospectively extract information on how individuals change and adapt their behavior during a pandemic. Moving forward, further research can focus on a more formal description of the relationship between the TE matrix and the age-mixing matrices used in spreading models, which may provide tools for model-free estimation of contact patterns during a pandemic.

\appendix
\supplementarysection

\section{KNN Information Estimators \label{app:knn}}

\begin{wrapfigure}[14]{l}{0.4\textwidth}
\vspace{-.9cm}
\begin{center}
\includegraphics[width=0.4\textwidth]{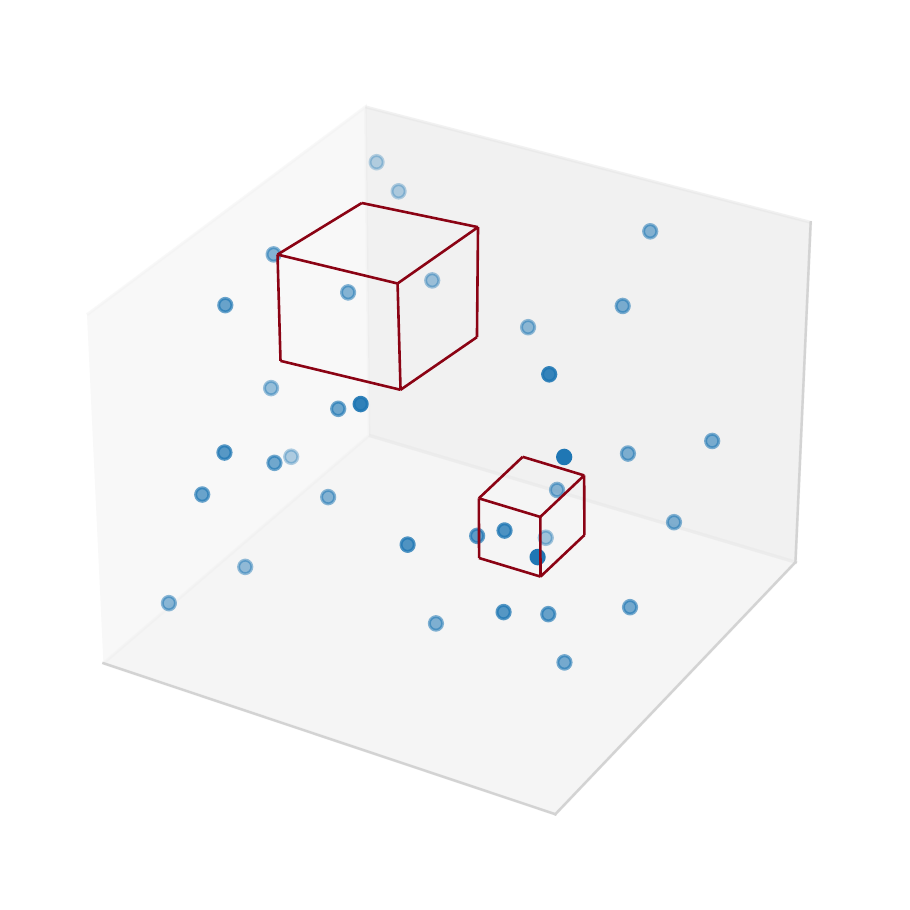}
\end{center}
\vspace{-.75cm}
\caption{\label{adaptativeness}Adaptativeness.}
\end{wrapfigure}

The k-nearest-neighbors (knn) are a class of estimators considered a nonparametric statistical method since can automatically adapt to any continuous underlying functions. Compared with kernel density estimators (kde), another nonparametric method, kNN method has some advantages. Firstly, kNN method usually require only one parameter tuning \cite{ksg}, see Fig.\ref{adaptativeness}. On the contrary, kde methods usually requires adjustment of bandwidth at each dimension separately, and thus the cost of parameter tuning is higher than kNN method, especially when dealing with high dimensional problems \cite{kde}.
Also, numerical experiments suggest that for the estimation of information theoretic functionals, kNN methods can usually outperform Kernel method \cite{thesis, outknn2, outknn1}. As a result, kNN methods are widely used for nonparametric statistical problems \cite{ksg}.

The knn estimator has as free parameter the number of nearest-neighbors $k$ in the joint space of $\mathcal{X} \times \mathcal{Y} \times \mathcal{Z}$ which determines the size of hyper-cubes, $\epsilon_i$, around each (high-dimensional) sample point $i$, Fig.\ref{adaptativeness}. Then $k_{Z,i}, k_{XZ,i}, k_{YZ,i}$ are  computed by counting the numbers of neighbors (points with distance strictly smaller than $\epsilon_i$ (including the reference point $i$) in the respective subspaces $\mathcal{Z}$, $\mathcal{X} \times \mathcal{Z}$, and $\mathcal{Y} \times \mathcal{Z}$. The hypercubes around each sample point are smaller where more samples are available.
\vspace{1.5cm}

\subsection{Estimation of Entropy and Mutual Information}

To get the CMI estimator, we first have to derive an entropy estimator since the former can be decomposed as a linear combination of the later. Indeed, in the continuous case, where we have a distribution density $p(x, y, z)$ of $(X,Y,Z)$, with the marginal densities $p(x,z), p(y,z)$, and $p(z)$, we have that the conditional mutual information is
\begin{equation}
I(X;Y|Z) = \int p(z)  \iint  p(x,y|z) \log
\frac{ p(x,y |z)}{p(x|z)\cdot p(y |z)} \,dx dy dz.   
\label{cmi_cont}
\end{equation}The term,
\begin{equation}
h(X) = - \int p(x) \log p(x) \,dx    
\end{equation}is called differential entropy of any random variable $X$ with density $p(x)$. Then, Eq.\ref{cmi_cont} can also be written in terms of differential entropies, 
\begin{equation}\label{app:cmi}
I(X;Y|Z) = h(X, Z) + h(Y, Z) - h(X, Y, Z) - h(Z).
\end{equation}If the condition $Z$ is irrelevant, i.e. $p(x, y, z) = p(x, y)p(z)$m CMI is equal to MI, 
\begin{equation}\label{app:mi}
I(X;Y) = h(X) + h(Y) - h(X, Y).
\end{equation}

For our purposes, the three time series $x_t$, $y_t$, and $z_t$ are considered as finite realizations of underlying stationary ergodic processes $X_t$, $Y_t$, and $Z_t$, respectively \cite{frenzelpompe}. Given $N$ points $\{x_i\}$ drawn from the pdf $p(x)$, the expression of Kozachenko-Leonenko estimator for the differential entropy is given by \cite{kl}:
\begin{equation}\label{app:ent}
\widehat{h}(X) = -\psi(k) +  \psi(N) + \text{ln}\, c_{d_x} + \frac{d_x}{N}
\sum_{i=1}^{N}\text{ln}\, \epsilon_i,   
\end{equation}in which $\psi$ is the digamma function defined as $\psi(x) = \Gamma'(x)/\Gamma(x)$,
\begin{equation}
\Gamma(x) = \int_{0}^{\infty}u^{x-1}e^{-u}du,
\end{equation}and $\epsilon_i$ is the distance from $x_i$ to its $k$-th nearest neighbor. The distance is defined as $d(x, x') =||x - x'||$ in which $||\cdot||$ can be any norm and $c_{d_x}$ represents the volume of the corresponding unit norm ball.

In order to obtain $I(X,Y|Z)$, we have to apply \ref{app:ent} in $\mathcal{X} \times \mathcal{Y} \times \mathcal{Z}$ \ref{app:cmi} adding and subtracting, respectively, this from estimates for $\widehat{h}(Z)$, and $\widehat{h}(X,Z)$, $\widehat{h}(Y,Z)$. For these terms, we could use Eq.\ref{app:ent} directly with the same $k$. But this would imply to use different distance scales in the joint and marginal spaces $\mathcal{X} \times \mathcal{Z}$, $\mathcal{Y} \times \mathcal{Z}$ and $\mathcal{Z}$. For any fixed $k$, the distance to the $k$-th neighbor in the joint space will be larger than the distances to the neighbors in the marginal spaces. Since the effect of the bias in Eq.\ref{app:ent} depends on the $k$-th neighbor distances, the biases in $\widehat{h}(X,Y,Z)$, $\widehat{h}(X,Z)$, $\widehat{h}(Y,Z)$ and $\widehat{h}(Z)$ would be different and would thus not cancel.

To deal with this this, we notice that in Eq.\ref{app:ent} we do not have to choose a fixed $k$ when
estimating the marginal entropies \cite{ksg, frenzelpompe}. Assume that there are altogether $k_{z_{i}}$ points within the vertical lines $||z-z_i||\leq\epsilon_{i}$, then $\epsilon_{i}$ is the distance to the $(k_{z_{i}}+1)$-st neighbor of $z_i$, and
\begin{equation}\label{app:est_ent}
\widehat{h}(z) = -\frac{1}{N}\sum_{i=1}^{N}\psi(k_{z_{i}}+1) +  \psi(N) + \text{ln}\, c_{d_z} + \frac{d_z}{N}
\sum_{i=1}^{N}\text{ln}\, \epsilon_i,   
\end{equation}
For the other subspaces this is not  exactly true, i.e., $\epsilon_{i}$ is not exactly equal to the distance to the $(k_{xz_{i}}+1)$-st neighbor, if $k_{xz_{i}}$ is defined as the number of points with $\max\{||x-x_i||, ||z-z_i||\}\leq\epsilon_{i}$ (similar to the $(k_{yz_{i}}+1)$-st neighbor). Nevertheless, we can consider Eq.\ref{app:est_ent} as a good approximation (indeed, it becomes exact when $(k_{xz_{i}}+1)$ or $(k_{yz_{i}}+1)$ $\rightarrow \infty$, and thus also when $N \rightarrow \infty$) for $\widehat{h}(X,Z)$ and $\widehat{h}(Y,Z)$, if we replace everywhere $Z$ by $(X,Z)$ and $(Y,Z)$ in its right-hand side . If we do this, we have the estimator
\begin{equation}
\widehat{I}(X;Y|Z) =  \frac{1}{N} \sum_{n=1}^N
\left[ \psi(k_{xz_{i}}+1) + \psi(k_{yz_{i}}+1) - \psi(k_{z_{i}}+1) \right] - \psi(k)
\end{equation}

Note that, If we disregard the condition $Z=\emptyset$, we have that $\psi(k_{z_{i}}+1)=\psi(N)$, $k_{xz_{i}}=k_{x_{i}}$ and $k_{yz_{i}}=k_{y_{i}}$. This yields the estimator for MI, $I(X, Y)=I(X, Y|\emptyset)$, as derived in \cite{ksg}, 
\begin{equation}
\widehat{I}(X;Y) = \frac{1}{N} \sum_{n=1}^N
\left[ \psi(k_{x_{i}}+1) + \psi(k_{y_{i}}+1) \right] - \psi(N) - \psi(k)
\end{equation}

\section{The shuffling technique \label{app:shuff}}

We here adopted a well-known approach in non-parametric statistics, called the permutation test. Specifically, we did the following permutation test based on the null hypothesis that $T_{X \rightarrow Y} = 0$: first perform $M$ random (temporal) permutations of the time series $X_{(i)}$, leaving the rest of the data unchanged; then construct an empirical cumulative distribution, $F(x)$, of the estimated $T_{X_{(i)} \rightarrow Y}(k,l)$; then, given a prescribed significance level $\alpha$, the observed $T_{X \rightarrow Y} = c$ is declared significant (i.e., the null hypothesis is rejected at level $\alpha$) if $F(c) > \alpha$.


\section{Temporal Effective information \label{app:eff} } 

%





The idea of defining a causal effect as the effect of an intervention in the context of time series has already been discussed in literature \cite{Eichler2012} as well as theoretic-informational approaches to deal with Markov-chains-generated time series \cite{Hoel2016}. Here, we expand the later formalism to deal with non-Markovian/continuous time series.


First, let's fix some terminology. We will focus on a given space $\Omega$, i.e., the set of all possible occurrences. In this space, we can consider causes $c \in \Omega$ and effects $e \in \Omega$, where we assume causes $c$ to precede effects $e$, so that we also speak of a set of causes $\mathcal{C} \subseteq \Omega$ and of effects $\mathcal{E} \subseteq \Omega$. We can consider the space $\Omega$ to be a state-space and $c$ or $e$ as states in the case of Markov chains or even discrete time-series. The set of causes and effects is related via conditional probabilities $P(e|c)$, which specifies the probability of obtaining a candidate effect $e$, given that a candidate causes $c$ actually occurred.


Information-based measures of causation (interventionist formalism) involve background assumptions in order to apply them \cite{Hoel2016}. The effect distribution given that we have intervened in the cause :
\begin{equation}
P(e) = \sum_{c\in \mathcal{C}} P(\text{do}(c))P(e | \text{do}(c))    
\end{equation}elucidates that there is some assumption about the distribution $P(C)=\{p(\text{do}(c))\}_{c \in \mathcal{C}}$ to meaningfully talk about $P(e)$. This assumption is necessary because, unlike terms like $P(e | \text{do}(c))$ which can be stated in data, terms like $P(\text{do}(c))$ need to be explicitly defined (e.g., what is the distribution of the effects when $c$ did not occur?).

To deal with this we will use distributions of interventions \cite{Hoel2016}. A distribution of intervention is a probability distribution over possible interventions that a modeler or experimenter considers. That is, instead of considering a single $\text{do}(c)$ operator \cite{pearl}, it is a probability distribution over some applied set of them. The interventional distribution fixes $P(C)$, the probability of all causes \cite{hoel}. With this in hand, we can define an informational-based causation measure for discrete finite systems, the effective information ($EI$),
\begin{equation}
EI:=I(C;E)=\mathbb{E}_{p(C)}D_{KL}\big(p(E|c)||P(E)\big),   
\end{equation}where $P(E)= \sum_{e}P(e)$. As a measure of causation, $EI$ captures how effectively (deterministically and uniquely) causes produce effects in the system, and how selectively causes can be identified from effects.

\subsection{Causal Analysis across scales: the most effective information}

A causal system can be represented in different scales either at different coarse-grains or over different subsets. Each such scale (here meaning a coarse-grain or some subset) can be treated as a particular causal model. The full micro causal model of a system is its most fine-grained representation in space and time over all elements and states ($micro:=\{S_m\}$). However, systems can also be considered as many different macro causal models ($macro:=\{S_M\}$)), such as higher scales or over a subset of the state-space. The set of all possible causal models, $\{S\}$, is entirely fixed by the base $micro=\{S_m\}$. In technical terms this is known as supervenience: given the lowest scale of any system (the base), all the subsequent macro causal models of that system are fixed. 


Macro causal models are defined as a mapping: $\mathcal{M}:micro \mapsto macro$, which can be a mapping in space, time, or both. As they are similar mathematically, here we only examine temporal mappings. One universal feature of a macroscale, in space and/or time, is its reduced size: $S_M$ must always be of a smaller cardinality than $S_m$. By assessing $EI(S)$ over all coarse grains of $S_m$, one can ask at which level of $\{S\}$ causation reaches a maximum. This provides a causation-based methodology to get the most effective scale containing more information in the relationships between the objects, expressed in bits:
\begin{equation}
EI(macro) - EI(micro)
\label{macromicro}
\end{equation}
Thus, if $EI$ is maximal for a macro-level rather than the micro-level, the quantity above is positive and  we say that such a scale is more effective to describe the system. These principles allow us to investigate causal emergence through temporal groupings, which coarse grain micro time steps $t_i$ into macro time steps $T_i$.

A last comment is that to quantify $EI(macro)$, we have to deal with macro interventions. In causal analysis, a coarse-grained intervention is an average over a set of micro-interventions as we will clarify in the next section. 

\subsection{Transfer Effective Information for Time-Series Data}

In Hoel's approach for measuring $EI$ across scales, no intervention is assumed to be more likely than any other based on the maximum entropy principle \cite{jaynes}. Perturbing using $H_{\text{max}}$ means intervening on some system overall $n$ possible states with equal probability so that $(\text{do}(c),\, \forall c \in \mathcal{C})$. However, the measure can only be applied to Markovian systems because, in general, for non-Markovian/continuous systems the conditional probability distribution cannot be uniquely specified by any maximum entropy generative model \cite{seth}. A proposal to overcome this situation is instead of measuring information generated by probabilities from a hypothetical maximum entropy past state, utilizing strategies based on the empirical distribution of the past state. This gives an expression to compare the \textit{transfer} effective information across scales:
\begin{equation}
I_{\text{macro}}\big(\mathbf{X}_{t}^{(\tau)};\mathbf{Y}_{t}| \mathbf{Y}_{t}^{(\tau)}\big) - I_{\text{micro}}\big(\mathbf{X}_{t}^{(\tau)}; \mathbf{Y}_{t}|\mathbf{Y}_{t}^{(\tau)}\big).
\end{equation}Here, we are focusing on temporal coarse grainings over the past of sources and targets time series up to lag $\tau$, $\mathbf{X}_{t}^{(\tau)}$ and $\mathbf{Y}_{t}^{(\tau)}$, respectively. However, we emphasize that the idea is not restricted to it. Also, note that the $EI$'s from Eq.\ref{macromicro} became simply mutual information due to the use of empirical distribution rather than maximum entropic distributions.

Finally, note that we can formalize Eq.\ref{map} in the main text as a macro intervention with the following strategy:
\begin{equation}
\text{do}\big(X_{agg}^{(\tau)}=x_{agg}^{(\tau)}\big) := \sum_{T=1}^{\tau} \gamma_{t-T} \,\text{do}\big(X_{t-T}=x_{t-T}\big), \quad \sum_{T=1}^{\tau} \gamma_{t-T}=1.
\end{equation}In this sense, the transfer-like quantifier,
\begin{equation}
I(X_{agg}^{(\tau)}; Y_t|Y_{agg}^{(\tau)}),
\end{equation}can be interpreted as measuring the causal impact of the macro intervention $X_{agg}^{(\tau)}$ in $Y_t$ given the knowledge of the macro intervention $Y_{agg}^{(\tau)}$. 

We can measure the impact of this \textit{transfer effective information} by computing the transfer entropy of the past random vector in the micro-level and taking the difference. This procedure tells us if this macro-scale intervention gives a better informative analysis. 

\begin{figure}[h]
\begin{center}
\includegraphics[width=\textwidth]{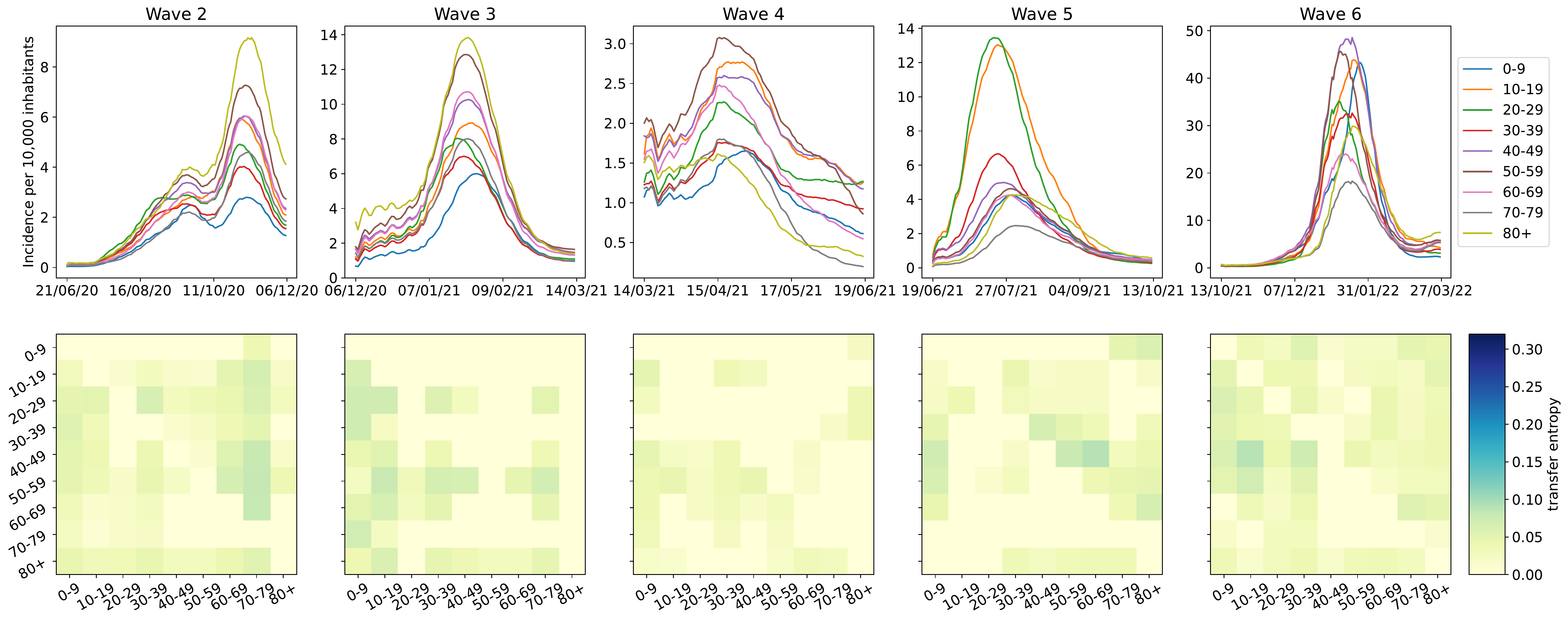}
\vspace{-.25cm}
\captionsetup{width=.95\textwidth}
\caption{Transfer entropy analysis of the COVID-19 pandemic in Spain without coarse-graining the past influence. Top row: incidence per age group in each official wave. The curves have been smoothed using a rolling average of 14 days to improve visualization. Note also that the scale of the $y$ axis changes in each wave. Bottom row: estimated TE across age groups in each wave.}
\label{fig:S_te_data}
\end{center}
\end{figure}

To confirm the argumentation we have that for all waves have $TE_{macro}/\log_2(N) > TE_{micro}/\log_2(n)$, see Table\ref{table}, where the division is responsable for a fair comparison among spaces with different dimension, in our case $N=3$ and $n=21$. Indeed, if we compare the TE matrices from Fig.\ref{fig:te_model} with Fig.\ref{fig:S_te_data} we can see how the patterns disappear and all of our analysis from Sec.\ref{results} is lost.

\begin{center}
\begin{table}[H]
\centering
\begin{tabular}{c c}
\toprule
$I(\mathbf{X}_{t}^{(\tau)}; \mathbf{Y}_t|\mathbf{Y}_{t}^{(\tau)})$ & $I(X_{agg}^{(\tau)}; Y_t|Y_{agg}^{(\tau)})$ \\
\midrule
0.45 & 1.18 \\
0.36 & 2.93 \\
0.21 & 3.19 \\
0.33 & 6.85 \\
0.50 & 3.68 \\
\bottomrule \\
\end{tabular}



\caption{The first and the second columns represent, respectively, the micro and macro level TE values for all age's groups. As we can see, all macro-level values beat the micro elucidating how this scale has a more effective analysis.}
\label{table}
\end{table}
\end{center}
\vspace{-1.cm}

\section{Further analysis \label{app:more} } 

\begin{figure}[H]
\begin{center}
\includegraphics[width=17cm]{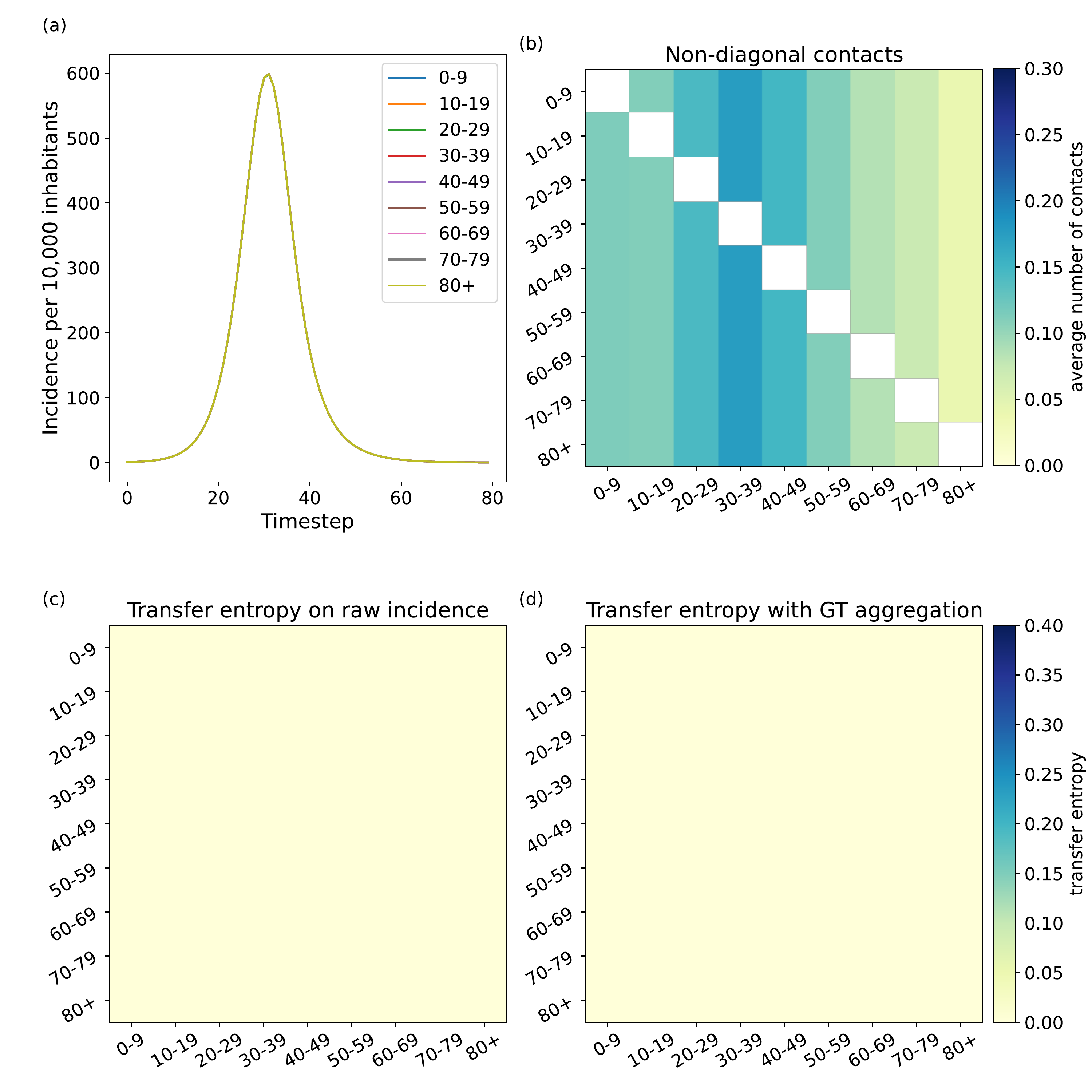}
\vspace{-.25cm}
\captionsetup{width=.95\textwidth}
\caption{Transfer entropy analysis of the SIR model. (a) Incidence per age group with $\beta=0.035$ and $\gamma = 1 / 5$. (b) Homogeneous mixing matrix ($M_{aa'} = N_{a'}/N$) without the diagonal contacts to highlight the interactions across groups. (c) Transfer entropy obtained using the raw incidence. (d) Transfer entropy obtained with the aggregated time series over the generation time.}
\label{fig:model_noage}
\end{center}
\end{figure}

In figure \ref{fig:model_noage}, we show the results obtained when we remove the age-mixing patterns, recovering the homogeneous mixing SIR model. Under this scenario all the incidence curves collapse on each other and the TE is 0 for all of them.

\let\clearpage\relax
\bibliographystyle{plain}

\end{document}